\documentstyle[12pt]{article}
\begin{flushright}
THES-TP 25/94\\
{}~DEM-NT  94-35
\end{flushright}

\begin{document}
\begin{center} {\Large\bf
Nonlinear extension of the u(2) algebra as the symmetry algebra of
the planar anisotropic quantum harmonic oscillator with rational ratio of
frequencies and ``pancake'' nuclei}

\bigskip\bigskip\bigskip

{Dennis Bonatsos$^{\ast,\dagger}$\footnote{e:mail
\begin{tabular}[t]{l}
bonat\@ cyclades.nrcps.ariadne-t.gr\\
bonat\@ nuclear.ect.unitn.it
\end{tabular}},
C.~Daskaloyannis$^{\dagger\dagger}$\footnote{
e:mail~daskaloyanni\@ olymp.ccf.auth.gr},
 P.~Kolokotronis$^\dagger$ and D. Lenis$^\dagger$}

\bigskip

{$^\ast$ European Centre for Theoretical Studies in Nuclear Physics
and Related Areas (ECT$^*$)}

{Strada delle Tabarelle 286, I-38050 Villazzano (Trento), Italy}

{$^\dagger$ Institute of Nuclear Physics, N.C.S.R.
``Demokritos''}

{GR-15310 Aghia Paraskevi, Attiki, Greece}

{$^{\dagger\dagger}$ Department of Physics, Aristotle University of
Thessaloniki}

{GR-54006 Thessaloniki, Greece }

\end{center}

\bigskip\bigskip\bigskip
\centerline{\bf Abstract}
\medskip
The symmetry algebra of the two-dimensional anisotropic quantum harmonic
oscillator with rational ratio of frequencies,
which is characterizing ``pancake''
nuclei,   is identified as a non-linear extension of
the u(2) algebra. The finite dimensional representation modules  of this
algebra are studied and the energy eigenvalues are determined  using
algebraic methods of general applicability to quantum superintegrable
systems. For labelling the degenerate states an ``angular momentum''
operator is introduced, the eigenvalues of which are roots of
appropriate generalized Hermite polynomials.  In the special case
 with frequency ratio 2:1 the resulting algebra is identified as the
 finite W algebra W$_3^{(2)}$.
\newpage

Quantum algebras \cite{Dri,Jim} (also called quantum groups) are nonlinear
deformations of the corresponding Lie algebras, to which they reduce when
the deformation parameter is set equal to unity. The interest in their
possible applications in physics was triggered by the introduction of the
q-deformed harmonic oscillator in 1989 \cite{Bie,Mac,Sun} as a tool
for providing a boson realization of the quantum algebra su$_q$(2), although
similar mathematical structures had already been known \cite{Ari,Kur}.
By now several kinds of generalized deformed oscillators
 (see \cite{Dask,Arik,Brz,PLB307,Mel} and references therein)
and generalized nonlinear deformed su(2) algebras
\cite{Kol,DQ1,DQ2,DQ3,LG,SunLi,Pan} have been introduced, finding applications
in a variety of physical problems.

On the other hand
the two-dimensional \cite{JH,Dem,Con,Cis,CLP,GKM} and three-dimensional
\cite{DZ,Mai,Ven,MV,RD,BCD} anisotropic harmonic oscillator have been the
subject of several investigations, both at the classical and
the quantum mechanical level. These oscillators are examples of
superintegrable systems \cite{Hie}. The special cases with frequency ratios
1:2 \cite{Holt,BW} and 1:3 \cite{FL} have also been considered. While
at the classical level it is clear that the su(N) or sp(2N,R) algebras
can be used for the description of the N-dimensional anisotropic oscillator,
the situation at the quantum level, even in the two-dimensional case,
is not as simple \cite{JH}.

In this letter we are going to prove that a generalized deformed u(2)
algebra is the symmetry algebra of the two-dimensional anisotropic
quantum harmonic oscillator, which is the oscillator describing the
single-particle level spectrum of ``pancake'' nuclei, i.e. of triaxially
deformed nuclei with $\omega_x >> \omega_y$, $\omega_z$ \cite{Rae}.

Let us consider the system described by the Hamiltonian:
\begin{equation}
H=\frac{1}{2}\left(
{p_x}^2 + {p_y}^2 + \frac{x^2}{m^2} + \frac{y^2}{n^2} \right),
\label{eq:Hamiltonian}
\end{equation}
where $m$ and $n$ are two natural numbers mutually prime ones, i.e.
their great common divisor is $\gcd (m,n)=1$.

We define the creation and annihilation
operators \cite{JH}
\begin{equation}
\begin{array}{ll}
a^\dagger=\frac{x/m - i p_x}{\sqrt{2}}, &
a =\frac{x/m + i p_x}{\sqrt{2}}, \\[0.24in]
b^\dagger=\frac{y/n - i p_y}{\sqrt{2}}, &
b=\frac{y/n + i p_y}{\sqrt{2}}.
\end{array}
\label{eq:operators}
\end{equation}
These operators satisfy the commutation relations:
\begin{equation}
\left[ a,a^\dagger \right] = \frac{1}{m},
\quad
\left[ b,b^\dagger \right] = \frac{1}{n},
\quad
\mbox{other commutators}=0.
\label{eq:commutators}
\end{equation}
Using eqs (\ref{eq:operators}) and (\ref{eq:commutators}) we can prove
by induction that:
$$
\begin{array}{ll}
\left[ a, \left( a^\dagger \right)^p \right] =
 \frac{p}{m} \left( a^\dagger \right)^{p-1} , &
\left[ b, \left( b^\dagger \right)^p \right] =
 \frac{p}{n} \left( b^\dagger \right)^{p-1} ,\\[0.24in]
\left[ a^\dagger, \left( a \right)^p \right] =
- \frac{p}{m} \left( a \right)^{p-1} , &
\left[ b^\dagger, \left( b \right)^p \right] =
- \frac{p}{n} \left( b \right)^{p-1} .
\end{array}
$$
Defining
$$
U=\frac{1}{2} \left\{ a, a^\dagger \right\}, \qquad
W=\frac{1}{2} \left\{ b, b^\dagger \right\},$$
one can easily prove that:
$$\begin{array}{ll}
\left[ U,
\left(a^\dagger \right)^p \right]= \frac{p}{m} \left(a^\dagger \right)^p, &
\left[ W,
\left(b^\dagger \right)^p \right]= \frac{p}{n} \left(b^\dagger \right)^p,
\\[0.24in]
\left[ U,
\left(a \right)^p \right]= - \frac{p}{m} \left(a \right)^p, &
\left[ W,
\left(b \right)^p \right]= - \frac{p}{n} \left(b \right)^p.
\end{array}
$$
Using the above properties we can define the enveloping algebra
generated by the operators:
\begin{equation}
\begin{array}{c}
S_+= \left(a^\dagger\right)^m \left(b\right)^n,\quad
S_-= \left(a\right)^m \left(b^\dagger\right)^n, \\[0.24in]
S_0= \frac{1}{2}\left( U -
W \right), \quad
H=U+W.
\end{array}
\label{eq:generators}
\end{equation}
These genarators satisfy the following relations:
\begin{equation}
\left[ S_0,S_\pm \right]=\pm S_\pm, \quad
\left[H,S_i\right]=0, \quad \mbox{for}\quad i=0,\pm,
\label{eq:SS}
\end{equation}
and
$$
S_+S_- =
\prod\limits_{k=1}^{m}\left( U - \frac{2k-1}{2m} \right)
\prod\limits_{\ell=1}^{n}\left( W + \frac{2\ell-1}{2n} \right),
$$
$$
S_-S_+ =
\prod\limits_{k=1}^{m}\left( U + \frac{2k-1}{2m} \right)
\prod\limits_{\ell=1}^{n}\left( W - \frac{2\ell-1}{2n} \right).
$$
The fact that the operators $S_i$, $i=0, \pm$ are integrals of motion
has been already realized in \cite{JH}.

The above relations mean that the harmonic oscillator of eq.
(\ref{eq:Hamiltonian})
is described by the enveloping algebra of the non-linear generalization
of the u(2) algebra formed by the generators $S_0$, $S_+$, $S_-$ and $H$,
satisfying the commutation relations of eq. (\ref{eq:SS}) and
\begin{equation}
\begin{array}{c}
\left[S_-,S_+\right] =
F_{m,n} (H,S_0+1) -  F_{m,n} (H,S_0),\\[0.24 in]
\mbox{where}\quad F_{m,n}(H,S_0)=
\prod\limits_{k=1}^{m}\left( H/2+S_0 - \frac{2k-1}{2m} \right)
\prod\limits_{\ell=1}^{n}\left( H/2-S_0 + \frac{2\ell-1}{2n} \right).
\end{array}
\label{eq:U2}
\end{equation}
This algebra is a non-linear generalization of the u(2) algebra,
of order $m+n-1$. In the case of $m/n=1/1$ this algebra is the
usual u(2) algebra, and the operators $S_0,S_\pm$ satisfy the
commutation relations of the ordinary su(2) algebra.

The finite dimensional representation modules
 of this algebra can be found
using the concept of the generalized deformed oscillator \cite{Dask},
in a method similar to the one used in \cite{BDK} for the study of
quantum superintegrable systems.
The operators:
\begin{equation}
{\cal A}^\dagger= S_+, \quad
{\cal A}= S_-, \quad
{\cal N}= S_0-u, \quad
u=\mbox{ constant},
\label{eq:alge-gen}
\end{equation}
where $u$ is a constant to be determined,
are the generators of a deformed oscillator algebra:
$$
\left[ {\cal N} , {\cal A}^\dagger \right] = {\cal A}^\dagger,
\quad
\left[ {\cal N} , {\cal A} \right] = -{\cal A},
\quad
{\cal A}^\dagger{\cal A} =\Phi( H, {\cal N} ),
\quad
{\cal A}{\cal A}^\dagger =\Phi( H, {\cal N}+1 ).
$$
 The structure function $\Phi$ of this algebra is determined by the function
$F_{m,n}$ in eq. (\ref{eq:U2}):
\begin{equation}
\begin{array}{l}
\Phi( H, {\cal N} )=
F_{m,n} (H,{\cal N} +u ) = \\
= \prod\limits_{k=1}^{m}\left( H/2+{\cal N} +u - \frac{2k-1}{2m} \right)
\prod\limits_{\ell=1}^{n}\left( H/2-{\cal N} -
u + \frac{2\ell-1}{2n} \right).
\end{array}
\label{eq:sf}
\end{equation}
The deformed oscillator corresponding to the structure function of eq.
 (\ref{eq:sf}) has an energy dependent Fock space of dimension $N+1$ if
\begin{equation}
\Phi(E,0)=0, \quad \Phi(E, N+1)=0, \quad
\Phi(E,k)>0, \quad \mbox{for} \quad k=1,2,\ldots,N.
\label{eq:equations}
\end{equation}
The Fock space is defined by:
\begin{equation}
H\vert E, k > =E \vert E, k >, \quad
{\cal N} \vert E, k >= k \vert E, k >,\quad
a\vert E, 0 >=0,
\end{equation}
\begin{equation}
{\cal A}^\dagger \vert E, k> =
\sqrt{\Phi(E,k+1)} \vert E, k+1>,
\quad
{\cal A} \vert E, k> =
\sqrt{\Phi(E,k)} \vert E, k-1>.
\end{equation}
The basis of the Fock space is given by:
$$
\vert E, k >= \frac{1}{\sqrt{[k]!}}
\left({\cal A}^\dagger\right)^k\vert E, 0 >,
\quad k=0,1,\ldots N,
$$
where the ``factorial''  $[k]!$ is defined by the recurrence relation:
$$
 [0]!=1, \quad [k]!=\Phi(E,k)[k-1]! \quad .
$$
Using the Fock basis we can find the matrix representation of the deformed
oscillator and then the matrix representation of the algebra of eqs.
(\ref{eq:SS}), (\ref{eq:U2}).
The solution of eqs (\ref{eq:equations}) implies the following pairs
of permitted values for the energy eigenvalue $E$ and the constant $u$:
\begin{equation}
E=N+\frac{2p-1}{2m}+\frac{2q-1}{2n} ,
\label{eq:E1}
\end{equation}
where $ p=1,2,\ldots,m$, $ q=1,2,\ldots,n$,
and
$$
u=\frac{1}{2}\left( \frac{2p-1}{2m}-\frac{2q-1}{2n} -N \right),
$$
the corresponding structure function being given by:
\begin{equation}
\begin{array}{l}
\Phi(E,x)=\Phi^{N}_{(p,q)}(x)=\\
=\prod\limits_{k=1}^{m}\left( x +
\displaystyle  \frac{2p-1}{2m}- \frac{2k-1}{2m} \right)
\prod\limits_{\ell=1}^{n}\left( N-x+
\displaystyle \frac{2q-1}{2n} +
\frac{2\ell-1}{2n}\right)\\
=\displaystyle\frac{1}{m^m n^n}
\displaystyle\frac{ \Gamma\left(mx+p\right) }{\Gamma\left(mx+p-m\right)}
\displaystyle
\frac{\Gamma\left( (N-x)n + q + n \right)}
{\Gamma\left( (N-x)n + q  \right)}.\end{array}
\label{eq:structure-function}
\end{equation}
In all these equations one has $N=0,1,2,\ldots$, while the dimensionality
of the representation is given by $N+1$. Eq. (\ref{eq:E1})
 means that there are $m\cdot n$ energy eigenvalues corresponding to each
 $N$ value, each eigenvalue having degeneracy $N+1$. (Later we shall see
that the degenerate states corresponding to the same eigenvalue can be
labelled by an ``angular momentum''.)
The energy formula can be corroborated by using the
corresponding Schr\"{o}dinger equation. For the Hamiltonian of eq.
(\ref{eq:Hamiltonian}) the eigenvalues of the Schr\"{o}dinger equation
are given by:
\begin{equation}
E=\frac{1}{m}\left(n_x+\frac{1}{2}\right)+
  \frac{1}{n}\left(n_y+\frac{1}{2}\right),
\label{eq:E2}
\end{equation}
where $n_x=0,1,\ldots$ and $n_y=0,1,\ldots$. Comparing eqs
(\ref{eq:E1}) and (\ref{eq:E2}) one concludes that:
$$N= \left[n_x/m\right]+\left[n_y/n\right],$$
where $[x]$ is the integer part of the number $x$, and
$$
p=\mbox{mod}(n_x,m)+1, \quad q=\mbox{mod}(n_y,n)+1.
$$

The eigenvectors of the Hamiltonian can be parametrized by the
dimensionality of the representation $N$, the numbers $p,q$,
and the number $k=0,1,\ldots,N$:

\begin{equation}
H\left\vert \begin{array}{c} N\\ (p,q) \end{array}, k \right>=
\left(N+\displaystyle
\frac{2p-1}{2m}+\frac{2q-1}{2n}
\right)\left\vert \begin{array}{c} N\\ (p,q) \end{array}, k \right>,
\label{eq:en-rep}
\end{equation}
\begin{equation}
S_0
\left\vert \begin{array}{c} N\\ (p,q) \end{array}, k \right>=
\left(
k+ \displaystyle
\frac{1}{2}
\left( \frac{2p-1}{2m}- \frac{2q-1}{2n} -N \right) \right)
\left\vert \begin{array}{c} N\\ (p,q) \end{array}, k \right>,
\label{eq:s0-rep}
\end{equation}
\begin{equation}
S_+\left\vert \begin{array}{c} N\\ (p,q) \end{array}, k \right>
= \sqrt{ \Phi^N_{(p,q)}(k+1)}
\left\vert \begin{array}{c} N\\ (p,q) \end{array}, k +1\right>,
\label{eq:sp-rep}
\end{equation}
\begin{equation}
S_-\left\vert \begin{array}{c} N\\ (p,q) \end{array}, k \right>
= \sqrt{ \Phi^N_{(p,q)}(k)}
\left\vert \begin{array}{c} N\\ (p,q) \end{array}, k -1\right>.
\label{eq:sm-rep}
\end{equation}

It is worth noticing that the operators $S_0,S_\pm$ do not
correspond to a generalization of the angular momentum,
$S_0$ being the operator corresponding to the Fradkin operator
$S_{xx}-S_{yy}$ \cite{Higgs,Leemon}. The corresponding ``angular
momentum'' is defined by:
\begin{equation}
L=-i\left(S_+-S_-\right).
\label{eq:angular-momentum}
\end{equation}
The ``angular momentum'' operator commutes with the Hamiltonian:
$$ \left[ H,L \right]=0. $$
Let $\vert \ell> $ be the eigenvector of the operator $L$ corresponding
to the eigenvalue $\ell$. The general form of this eigenvector
can be given by:
\begin{equation}
\vert \ell > = \sum\limits_{k=0}^N
\frac{i^k c_k}{\sqrt{[k]!}}
\left\vert \begin{array}{c} N\\ (p,q) \end{array}, k \right>.
\end{equation}

In order to find the eigenvalues of $L$  and the coefficients $c_k$
we use the Lanczos algorithm \cite{Lan}, as formulated
 in \cite{Flo}. From eqs (\ref{eq:sp-rep}) and (\ref{eq:sm-rep}) we find
\begin{equation}
\begin{array}{l}
L\vert \ell > =\ell \vert  \ell >=
\ell\sum\limits_{k=0}^N
\frac{i^k c_k}{\sqrt{[k]!}}
\left\vert \begin{array}{c} N\\ (p,q) \end{array}, k \right>=
\\
=\frac{1}{i}
\sum\limits_{k=0}^{N-1}
\frac{i^k c_k \sqrt{\Phi^N_{(p,q)}(k+1)}}{\sqrt{[k]!}}
\left\vert \begin{array}{c} N\\ (p,q) \end{array}, k+1 \right>-
\frac{1}{i}
\sum\limits_{k=1}^{N}
\frac{i^k c_k \sqrt{\Phi^N_{(p,q)}(k)}}{\sqrt{[k]!}}
\left\vert \begin{array}{c} N\\ (p,q) \end{array}, k-1 \right>
\end{array}
\end{equation} From this equation we find that:
$$c_k=  (-1)^k    2^{-k/2}  H_k (\ell /\sqrt{2} ), $$
where the function $H_k(x)$ is a generalization of the
``Hermite'' polynomials (see also \cite{PLB331,KZ}),
satisfying the recurrence relations:
$$
H_{-1}(x)=0, \quad H_0(x)=1,
$$
$$
H_{k+1}(x)= 2 x H_k(x) - 2\Phi^N_{(p,q)}(k) H_{k-1}(x),
$$
and the ``angular momentum'' eigenvalues $\ell$ are the roots of the polynomial
equation:
$$
H_{N+1}(\ell/\sqrt{2}) = 0.
$$
Therefore for a given value of $N$ there are $N+1$ ``angular momentum''
eigenvalues $\ell$, symmetric around zero
 (i.e. if $\ell$ is an ``angular momentum'' eigenvalue,
then $-\ell$ is also an ``angular momentum'' eigenvalue).
In the case of the symmetric harmonic oscillator ($m/n=1/1$)
these eigenvalues are uniformly distributed and differ by 2. In the
general case the ``angular momentum'' eigenvalues are non-uniformly
distributed. For small values of $N$ analytical formulae for the ``angular
momentum'' eigenvalues can be found \cite{PLB331}. Remember that to each
value of $N$ correspond $m\cdot n$ energy levels, each with degeneracy $N+1$.

 For the special case $m = 1$, $n=2$  it should be noticed that the nonlinear
 deformed algebra received here coincides with the finite W algebra
 W$_3^{(2)}$ \cite{Tj1,Tj2,Tj3}. The commutation relations of the
W$_3^{(2)}$ algebra are
$$ [H_W, E_W]= 2 E_W, \qquad [H_W, F_W]= -2 F_W, \qquad [E_W,F_W]= H^2_W
+ C_W, $$
$$ [C_W,E_W]=[C_W,F_W]=[C_W,H_W]=0,$$
while in the $m=1$, $n=2$ case one has the relations
$$ [{\cal N}, {\cal A}^{\dagger}]= {\cal A}^{\dagger}, \qquad
[{\cal N}, {\cal A}]= -{\cal A}, \qquad
[{\cal A},{\cal A}^{\dagger}]= 3 S_0^2 -{H^2\over 4} - H S_0 +{3 \over 16},$$
$$[H, {\cal A}^{\dagger}]= [H, {\cal A}]= [H, S_0] =0,$$
with $S_0= {\cal N}+u$ (where $u$ a constant).
It is easy to see that the two sets of commutation relations are
equivalent by making the identifications
$$ F_W= \sigma {\cal A}^{\dagger}, \qquad E_W= \rho {\cal A}, \qquad
H_W= -2 S_0 + k H, \qquad C_W= f(H), $$
with $$ \rho \sigma = {4\over 3}, \qquad k={1\over 3}, \qquad f(H)=-{4\over 9}
H^2 +{1\over 4}.$$

In conclusion,
the two-dimensional anisotropic quantum harmonic oscillator with rational ratio
of frequencies equal to $m/n$, which characterizes the single-particle
level spectrum of ``pancake'' nuclei,
is described dynamically by a non-linear extension of the
u(2) Lie algebra, the order of this algebra being $m+n-1$.
The representation modules of this algebra can be generated
by using the deformed oscillator algebra. The energy eigenvalues are
calculated by the requirement of the existence of finite dimensional
representation modules. An ``angular momentum'' operator useful for
labelling degenerate states has also been constructed.
 In the special case
 of $m:n=1:2$ the resulting algebra has been identified as the finite W
 algebra W$_3^{(2)}$.

The extension of the present method to the three-dimensional anisotropic
quantum harmonic oscillator is already receiving attention, since it is
of clear interest in the study of the symmetries underlying the
structure of superdeformed and hyperdeformed nuclei \cite{Mot}.

 The authors are grateful to Tjark Tjin for pointing out the
 correspondence between the $m:n =1:2$ case and the W$_3^{(2)}$ algebra.
One of the authors (DB) has been supported by the EU under contract
ERBCHBGCT930467. This project has also been partially supported by the  Greek
Secretariat of Research and Technology under contract PENED 340/91.

\vfill\eject


\begin{thebibliography}{99}

\bibitem{Dri}
V. G. Drinfeld, in {\it Proceedings of the International
Congress of Mathematicians}, ed. A. M. Gleason
(American Mathematical Society, Providence, RI, 1986) p. 798.

\bibitem{Jim}
M. Jimbo, Lett. Math. Phys. 11 (1986) 247.

\bibitem{Bie}
L. C. Biedenharn, J. Phys. A 22 (1989) L873.

\bibitem{Mac}
A. J. Macfarlane, J. Phys. A 22 (1989) 4581.

\bibitem{Sun}
C. P. Sun and H. C. Fu, J. Phys. A 22 (1989) L983.

\bibitem{Ari}
M. Arik and D. D. Coon, J. Math. Phys. 17 (1976) 524.

\bibitem{Kur}
V. V. Kuryshkin, Annales de la Fondation Louis de Broglie 5 (1980) 111.

\bibitem{Dask}
C. Daskaloyannis, J. Phys. A 24 (1991) L789.

\bibitem{Arik}
M. Arik, E. Demircan, T. Turgut, L. Ekinci and M. Mungan, Z. Phys. C
55 (1992) 89.

\bibitem{Brz}
T. Brzezi\'nski, I. L. Egusquiza and A. J. Macfarlane, Phys. Lett. B 311
(1993) 202.

\bibitem{PLB307}
D. Bonatsos and C. Daskaloyannis, Phys. Lett. B 307 (1993) 100.

\bibitem{Mel}
S. Meljanac, M. Milekovic and S. Pallua, Phys. Lett. B 328 (1994) 55.

\bibitem{Kol}
D. Bonatsos, C. Daskaloyannis and P. Kolokotronis, J. Phys. A 26 (1993)
L871.

\bibitem{DQ1}
C. Delbecq and C. Quesne, J. Phys. A 26 (1993) L127.

\bibitem{DQ2}
C. Delbecq and C. Quesne, Phys. Lett. B 300 (1993) 227.

\bibitem{DQ3}
C. Delbecq and C. Quesne, Mod. Phys. Lett. A 8 (1993) 961.

\bibitem{LG}
A. Ludu and R. K. Gupta, J. Math. Phys. 34 (1993) 5367.

\bibitem{SunLi}
C. P. Sun and W. Li, Commun. Theor. Phys. 19 (1993) 191.

\bibitem{Pan}
F. Pan, J. Math. Phys. 35 (1994) 5065.

\bibitem{JH}
J. M. Jauch and E. L. Hill, Phys. Rev. 57 (1940) 641.

\bibitem{Dem}
Yu. N. Demkov, Soviet Phys. JETP 17 (1963) 1349.

\bibitem{Con}
G. Contopoulos, Z. Astrophys. 49 (1960) 273; Astrophys. J. 138 (1963) 1297.

\bibitem{Cis}
A. Cisneros and H. V. McIntosh, J. Math. Phys. 11 (1970) 870.

\bibitem{CLP}
O. Casta\~nos and R. L\'opez-Pe\~na, J. Phys. A 25 (1992) 6685.

\bibitem{GKM}
A. Ghosh, A. Kundu and P. Mitra, Saha Institute preprint SINP/TNP/92-5
(1992).

\bibitem{DZ}
F. Duimio and G. Zambotti, Nuovo Cimento 43 (1966) 1203.

\bibitem{Mai}
G. Maiella, Nuovo Cimento 52 (1967) 1004.

\bibitem{Ven}
I. Vendramin, Nuovo Cimento 54 (1968) 190.

\bibitem{MV}
G. Maiella and G. Vilasi, Lettere Nuovo Cimento 1 (1969) 57.

\bibitem{RD}
G. Rosensteel and J. P. Draayer, J. Phys. A 22 (1989) 1323.

\bibitem{BCD}
D. Bhaumik, A. Chatterjee and B. Dutta-Roy, J. Phys. A 27 (1994) 1401.

\bibitem{Hie}
J. Hietarinta, Phys. Rep. 147 (1987) 87.

\bibitem{Holt}
C. R. Holt, J. Math. Phys. 23 (1982) 1037.

\bibitem{BW}
C. P. Boyer and K. B. Wolf, J. Math. Phys. 16 (1975) 2215.

\bibitem{FL}
A. S. Fokas and P. A. Lagerstrom, J. Math. Anal. Appl. 74 (1980) 325.

\bibitem{Rae}
W. D. M. Rae, Int. J. Mod. Phys. A 3 (1988) 1343.

\bibitem{BDK}
D. Bonatsos, C. Daskaloyannis and K. Kokkotas, Phys. Rev. A 48 (1993) R3407.

\bibitem{Higgs}
P. W. Higgs, J. Phys. A 12 (1979) 309.

\bibitem{Leemon}
H. I. Leemon, J. Phys. A 12 (1979) 489.

\bibitem{Lan}
C. Lanczos, J. Res. Natl. Bur. Stand. 45 (1950) 255.

\bibitem{Flo}
R. Floreanini, J. LeTourneux and L. Vinet, Ann. Phys. 226 (1993) 331.

\bibitem{PLB331}
D. Bonatsos, C. Daskaloyannis, D. Ellinas and A. Faessler, Phys. Lett. B
331 (1994) 150.

\bibitem{KZ}
A. A. Kehagias and G. Zoupanos, Z. Phys. C 62 (1994) 121.

\bibitem{Tj1}
T. Tjin, Phys. Lett. B 292 (1992) 60.

\bibitem{Tj2}
J. de Boer and T. Tjin, Commun. Math. Phys. 158 (1993) 485.

\bibitem{Tj3}
T. Tjin, private communication.

\bibitem{Mot}
B. Mottelson, Nucl. Phys. A 522 (1991) 1c.

\end{thebibliography}
\end{document}